\documentclass{PoS}

\usepackage{amsmath,amssymb}

\newcommand{\be}{\begin{equation}}
\newcommand{\ee}{\end{equation}}
\newcommand{\bea}{\begin{eqnarray}}
\newcommand{\eea}{\end{eqnarray}}

\newcommand{\rk}{\right)}
\newcommand{\lk}{\left(}

\newcommand*{\dbar}{\mathop{\mathrm{d}\mkern-5.5mu\mathchar'26}}

\newcommand{\vB}{\vec{B}}

\newcommand{\vA}{\vec{A}}
\newcommand{\vx}{\vec{x}}

\newcommand{\vk}{\vec{k}}

\newcommand{\vq}{\vec{q}}
\newcommand{\vp}{\vec{p}}

\DeclareMathOperator{\Tr}{Tr} 
\DeclareMathOperator{\Det}{Det}

\title{Hamilton Approach to QCD in Coulomb Gauge: Finite Temperatures and Chiral Symmetry Breaking}

\ShortTitle{Hamilton Approach to QCD in Coulomb Gauge}

\author{\speaker{Hugo Reinhardt}\\
Universit\"at T\"ubingen, Institut f\"ur Theoretische Physik\\
Auf der Morgenstelle 14, 72076 T\"ubingen, Germany\\
E-mail: \email{hugo.reinhardt@uni-tuebingen.de}}

\author{Davide R.~Campagnari, Jan Heffner and Markus Pak\\
Universit\"at T\"ubingen, Institut f\"ur Theoretische Physik\\
Auf der Morgenstelle 14, 72076 T\"ubingen, Germany
}

\abstract{I will review results obtained recently within the Hamilton approach to QCD
in Coulomb gauge. The focus will be on finite-temperature Yang--Mills theory and chiral
symmetry breaking in QCD.}

\FullConference{International Workshop on QCD Green's Functions, Confinement and Phenomenology\\
		September 7--11, 2011\\
		ECT Trento, Italy}

\begin{document}

\section{Introduction}

In this talk I will give an overview over recent results obtained within the Hamiltonian
approach to QCD in Coulomb gauge. I will focus on the extension of this approach to finite
temperatures and to the inclusion of quarks. My main interest here will be the
finite-temperature deconfinement phase transition and the spontaneous breaking of chiral symmetry. 

Going to Weyl gauge $A_0 = 0$ canonical quantization of Yang--Mills theory yields the following Hamiltonian
\be
\label{1} 
H = \frac{1}{2} \int \lk \vec{\Pi}^2 + \vB^2 \rk .
\ee
Here $\vec{\Pi} = \delta / i \delta \vA$ is the canonical momentum operator and $\vB$ is
the non-Abelian magnetic field. The Yang--Mills Hamiltonian (\ref{1}) is invariant under
spatial (time-independent) gauge transformations. A quantity of central interest in quantum
field theory is the vacuum wave functional, by means of which all correlation functions
can be evaluated. This quantity is obtained by solving the Schr\"odinger equation
\be
\label{2}
H \psi = E \psi
\ee
for the lowest energy eigenstate. There have been various attempts to solve the Yang--Mills
Schr\"o\-din\-ger equation (\ref{2}) directly for gauge invariant wave functionals, in
particular in $D = 2 + 1$, see Ref.~\cite{Greensite:2011pj} and references therein. One
can give arguments that the Yang--Mills vacuum wave functional (in $D = 3 + 1$) can be
approximated in the low-energy regime by \cite{Greensite:1979yn} 
\be
\label{3}
\psi [A] = \exp \left[ - \frac{1}{2} \int F^2_{ij} \right] .
\ee
So far one has not succeeded in determining the Yang--Mills vacuum wave functional in a
gauge invariant way. A much more convenient way is to fix the gauge and for the purpose
of Hamiltonian Yang--Mills theory Coulomb gauge is, in particular, convenient. The prize
one pays is that the Hamiltonian is more complicated. In Coulomb gauge the Yang--Mills
Hamiltonian is given by \cite{Christ:1980ku}
\be
\label{4}
H = \frac{1}{2} \int \lk J^{- 1} \vec{\Pi}^\perp J \vec{\Pi}^\perp + \vB^2 \rk + H_C \, ,
\ee
where
%\be
%\label{5}
$J = \Det (- \hat{D} \partial)$
%\ee
is the Faddeev--Popov determinant with $\hat{D} = \partial + g \hat{A}$, $\hat{A}^{ab} = f^{acb} A^c$
being the covariant derivative in the adjoint representation. Furthermore, 
\be
\label{6}
H_C = \frac{1}{2} \int J^{- 1} \vec{\Pi}^{\parallel} J \vec{\Pi}^{\parallel} = \frac{g^2}{2} \int J^{- 1} \rho (- \hat{D} \partial)^{- 1} (- \partial^2)
(- \hat{D} \partial)^{- 1} J \rho
\ee
is the Coulomb Hamiltonian, which arises from solving Gauss's law (which is a constraint
to the wave functional to guarantee gauge invariance) for the longitudinal momentum operator
$\vec{\Pi}^{\parallel}$. In Eq.~\eqref{6}, $\rho = \rho_{gl} + \rho_m$ is the total color
charge, which contains beside the charge of the Yang--Mills field $\rho_{gl} = - \hat{\vA} \vec{\Pi}$
also the charge of the matter fields $\rho_m$. By resolving Gauss's law gauge invariance
has been fully taken into account and in Coulomb gauge each functional of the transverse
gluon field $A^\perp$, $\partial A^\perp = 0$ is, in principle, a physical wave functional.
Note that in the canonical quantization of Yang--Mills theory the gauge field figures as
the coordinate, and the transition to Coulomb gauge implies a transition to curvilinear
coordinates. Accordingly, the kinetic piece of the Yang--Mills Hamiltonian [the first
term in Eq.~(\ref{4})] resembles the Laplacian in curvilinear coordinates. In the scalar
product of the Yang--Mills wave functionals the transition to Coulomb gauge can be
accomplished by using the standard Faddeev--Popov method, which introduces the Faddeev--Popov
determinant also in the integration measure
\be
\label{7}
\langle \phi | \dots | \psi \rangle = \int DA \: J (A) \, \phi^* (A) \dots \psi (A) .
\ee

%%%%%%%%%%%%%%%%%%%%%%%%%%%%%%%%%%%%%%%%%%%%%%%%%%%%%%%%%%%%%%%%%%%%%%%%%%%%%%%%%%%%%%%%%
%%%%%%%%%%%%%%%%%%%%%%%%%%%%%%%%%%%%%%%%%%%%%%%%%%%%%%%%%%%%%%%%%%%%%%%%%%%%%%%%%%%%%%%%%

\section{Zero-temperature Yang--Mills theory}

One can solve the Yang--Mills Schr\"odinger equation in perturbation theory by expanding
the Hamiltonian and the wave functional in powers of the coupling constant $g$, applying
standard Rayleigh--Schr\"odinger perturbation theory \cite{Campagnari:2009km}. From the
Coulomb term (\ref{6}), which is order $g^2$, one can extract the running coupling constant
and obtains the same result as in ordinary covariant perturbation theory within the
functional integral formulation. We are interested here, however, in a non-perturbative
solution of the Yang--Mills Schr\"odinger equation and for this purpose we exploit the
variational approach using Gaussian type wave functionals. The first variational calculations
in Coulomb gauge were performed in Ref.~\cite{Schutte:1985sd} and later on in Ref. \cite{Szczepaniak:2001rg}.
Our approach \cite{Feuchter:2004mk} differs from previous variational calculations in
Coulomb gauge in the ansatz of the vacuum wave functional, in the treatment of the Faddeev--Popov
determinant (treated fully in our approach and at least partially neglected in previous
approaches) and in the renormalization, for more details see Sect.~IID of Ref.~\cite{Greensite:2011pj}.

\begin{figure}[t]
\centering
\begin{minipage}[t]{.45\linewidth}
\includegraphics[width=\linewidth]{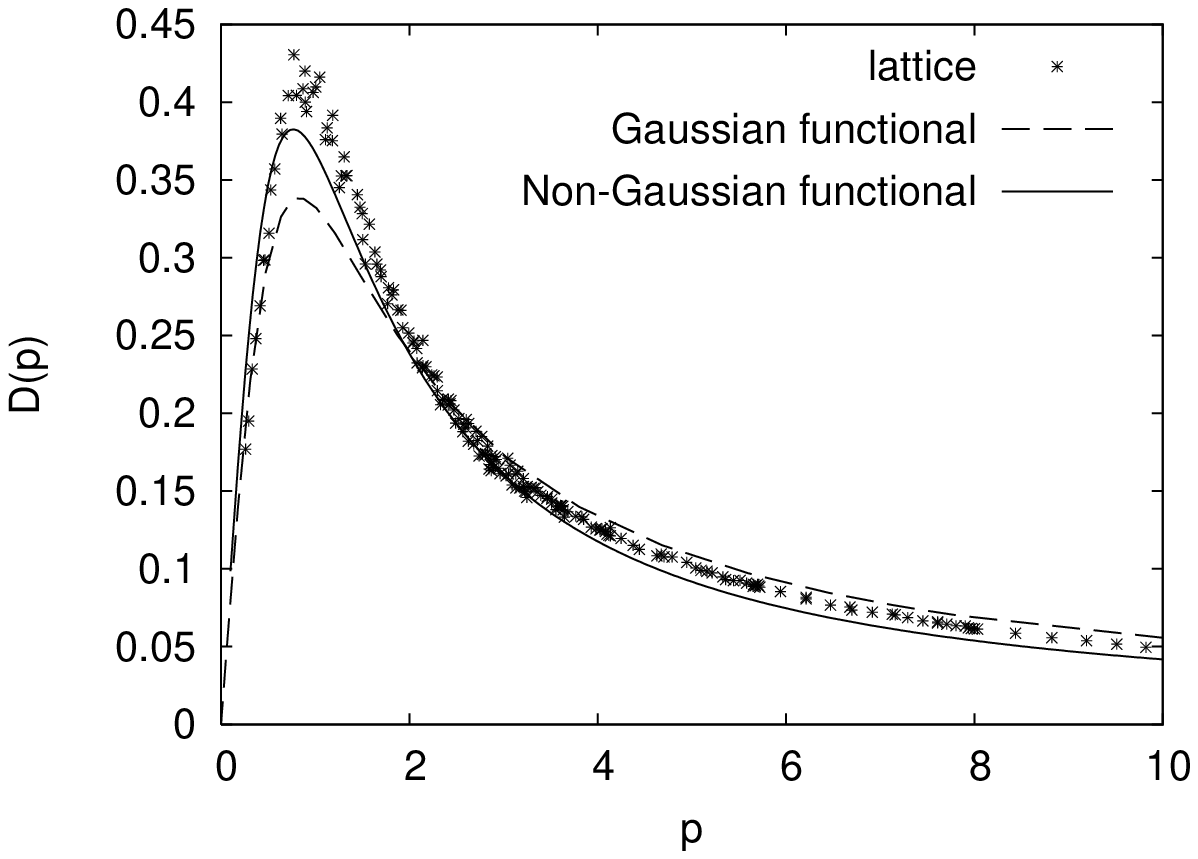}
\caption{\label{figure2}Comparison of the gluon propagator $1/(2\omega)$ with Gaussian
\cite{Epple:2006hv} (dashed line) and non-Gaussian \cite{Campagnari:2010wc} (full line)
functional to the lattice data \cite{Burgio:2008jr}.}
\end{minipage}
\hfil
\begin{minipage}[t]{.45\linewidth}
\includegraphics[width=\linewidth]{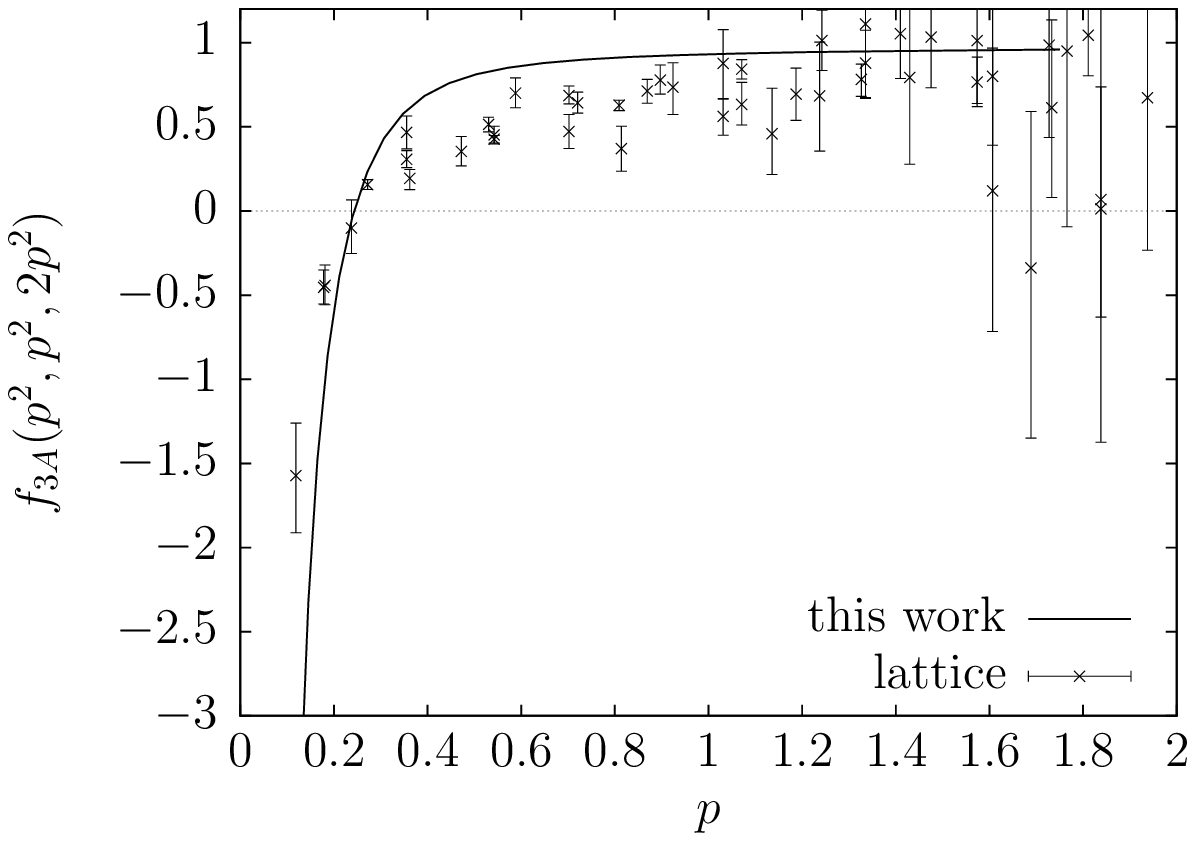}
\caption{\label{figure3}Form factor of the three-gluon vertex for orthogonal momenta
and comparison to lattice data for the 3-dimensional Landau-gauge vertex \cite{R11}.}
\end{minipage}
\end{figure}

The variational approach developed in T\"ubingen uses the trial ansatz for the vacuum
wave functional \cite{Feuchter:2004mk}
\be
\label{8}
\psi (A) = \frac{1}{\sqrt{\Det (- \hat{D} \partial)}} \exp \left[ -\frac{1}{2} \int A \omega A \right] ,
\ee
which contains besides the exponential the inverse square root of the Fadeev--Popov determined.
This has the advantage that in expectation values the Faddeev--Popov determinant in the
integration measure, see Eq.~(\ref{7}), is cancelled. Furthermore, the static gluon
propagator is with this wave functional given by the inverse of the variational kernel $\omega$
\be
\label{9}
\langle A A \rangle = (2 \omega)^{- 1} ,
\ee
which shows that $\omega$ has the meaning of the gluon energy. Minimizing the vacuum
expectation value of the Hamiltonian, $\langle \psi | H | \psi \rangle \to \mathrm{min}$,
one finds \cite{Epple:2006hv} for the gluon energy $\omega$ the result shown in
Fig.~\ref{figure2} (dashed line).
At large momenta it behaves like the photon energy $\omega (k\to\infty) \sim k$ and
is infrared divergent $\omega (k\to0) \sim 1/k$. This, of course, is the manifestation
of the confinement of gluons. Figure \ref{figure2} shows the lattice results \cite{Burgio:2008jr}
for the gluon propagator (\ref{9}), which can be fitted by Gribov's formula.
\be
\label{12}
\omega (k) = \sqrt{k^2 + \frac{M^4}{k^2}}
\ee
with the effective mass $M \approx 880$ MeV. Also shown in this figure is the result
of the variational calculation \cite{Epple:2006hv}: as can be seen the gluon propagator
obtained from the variational calculation agrees quite well with the lattice data in the
infrared and the ultraviolet, while there is some missing strength in the mid-momentum
regime around 1 GeV. This can be traced back to the absence of the gluon loop, which
escapes the variational calculation with the Gaussian type of ansatz (\ref{8}). In
Ref.~\cite{Campagnari:2010wc} the variational approach was extended to non-Gaussian wave
functionals including up to quartic terms in the gauge field
\be
\label{13}
|\psi [A]|^2 = \exp (- S [A]) , \qquad
S [A] = \int \omega A^2 + \frac{1}{3!} \int \gamma^{(3)} A^3 + \frac{1}{4!} \int \gamma^{(4)} A^4 .
\ee
To capture the gluon loop in the variational calculation one needs to include at least
the three-gluon term $\gamma^{(3)}$ in the exponent of the wave functional. Then one finds
the static gluon propagator shown in Fig.~\ref{figure2} (full line), which gives a
substantial improvement compared to the propagator obtained with the Gaussian trial wave
functional. Figure \ref{figure3} shows the result for the three-gluon vertex.
Also shown are the lattice result for the three-gluon vertex calculated in 3-dimensional
Landau gauge Yang--Mills theory \cite{R11}. This theory corresponds to the use of the
approximate wave functional (\ref{3}) in $3 + 1$ dimensional Yang--Mills theory in
the Hamiltonian approach. Since the expression (\ref{3}) represents a
good approximation to the true Yang--Mills wave functional in the low-momentum regime (see Ref.~\cite{Quandt:2010yq})
we expect the 3-dimensional Landau gauge lattice result to agree well with the static
three-gluon vertex in $D = 3 + 1$ Yang--Mills theory. Indeed we find a quite reasonable
agreement between the result of the variational calculation and the lattice data in the
infrared.

\begin{figure}[t]
\centering
\includegraphics[width=0.45\linewidth]{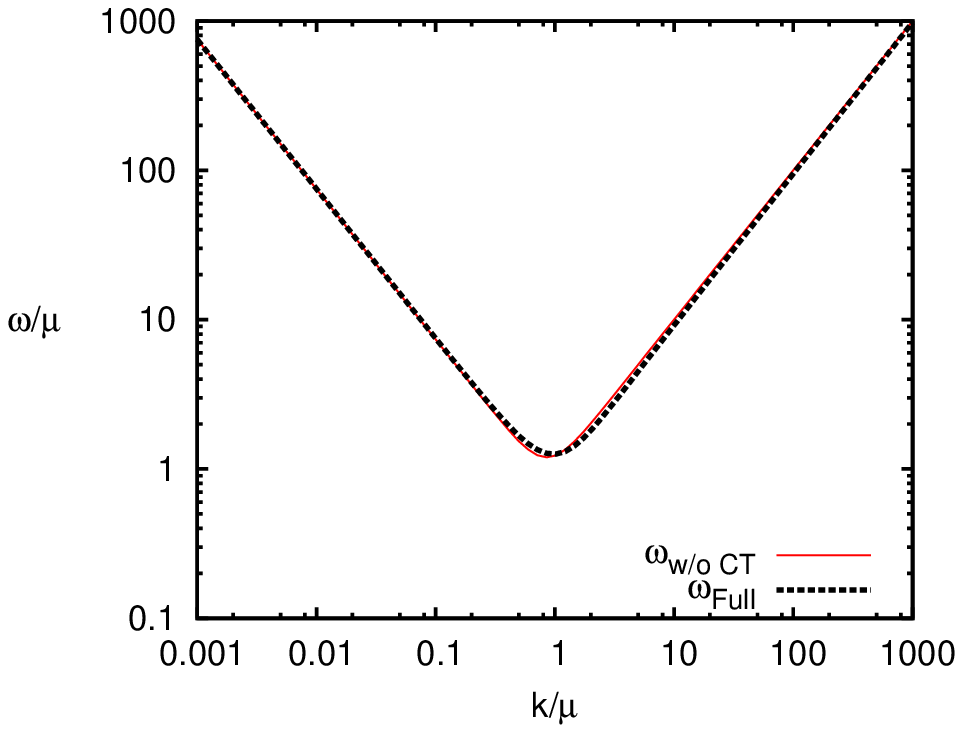}\hfil
\includegraphics[width=0.45\linewidth]{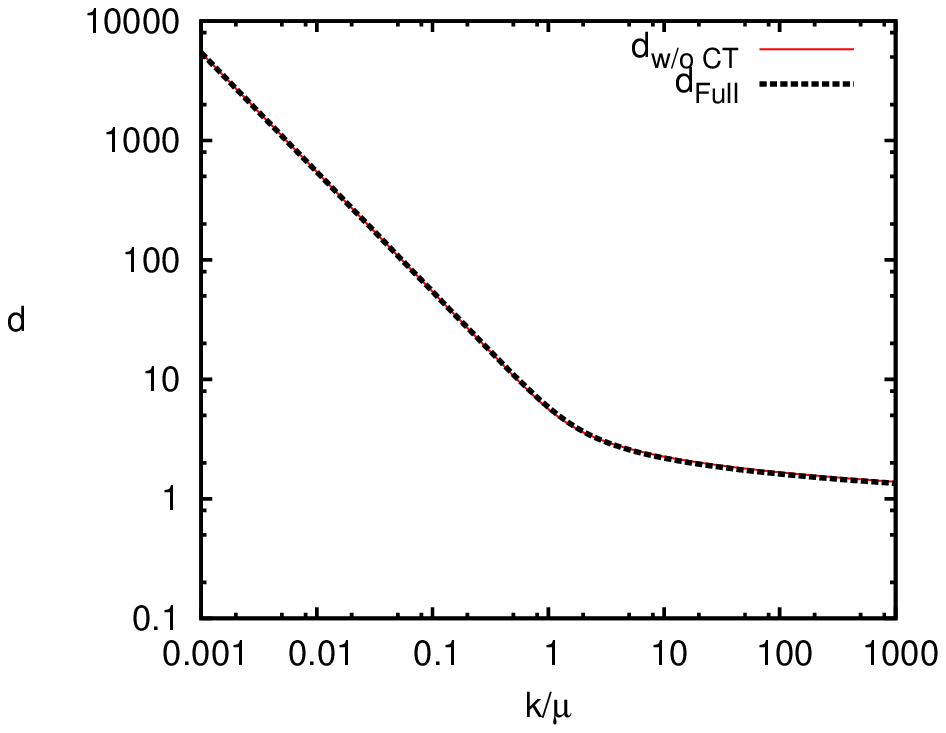}
\caption{\label{figure4}The gluon energy (left panel) and the ghost form factor (right panel) with and without the Coulomb term.}
\end{figure}

The above represented solution correspond to the so-called critical (or scaling) solution,
where the horizon condition $d^{- 1} (k = 0) = 0$ was assumed for the ghost form factor
$d (k)$ defined by the ghost propagator
\be
\label{16}
\langle (- \hat{D} \partial)^{- 1} \rangle  =  d (k) / k^2 .
\ee
Figure \ref{figure4} shows the result of the variational calculation for the gluon energy
(shown already in Fig.~\ref{figure2}) and the ghost form factor of the full variational
calculation and the one with the Coulomb term of the Hamiltonian, Eq.~(\ref{6}), excluded.
One observes that the effect of the Coulomb term is very small. We will therefore neglect
this term in subsequent numerical calculations. 

%%%%%%%%%%%%%%%%%%%%%%%%%%%%%%%%%%%%%%%%%%%%%%%%%%%%%%%%%%%%%%%%%%%%%%%%%%%%%%%%%%%%%%%%%
%%%%%%%%%%%%%%%%%%%%%%%%%%%%%%%%%%%%%%%%%%%%%%%%%%%%%%%%%%%%%%%%%%%%%%%%%%%%%%%%%%%%%%%%%

\section{Finite-temperature Yang--Mills theory}

To extend Yang--Mills theory to finite temperatures \cite{Reinhardt:2011hq, R14} we
consider the grand canonical ensemble with vanishing chemical potential, which is
defined by the density matrix
\be
\label{18}
D = \exp \bigl[- H/(k_{\textsc{b}} T) \bigr] ,
\ee
where $k_{\textsc{b}}$ is Boltzmann constant. For the evaluation of the thermal averages 
\be
\label{19}
\langle \dots \rangle_T = \frac{\Tr (D \dots)}{\Tr D}
\ee
we need a complete basis of the gluonic Fock space, which we choose
in analogy to the ground state wave functional (\ref{8}) in the form
\be
\label{20}
| \tilde{k} \rangle = \frac{1}{\sqrt{\Det (- \hat{D} \partial)}} | k \rangle ,
\ee
where the $| k \rangle$ form a complete set of states of the gluonic Fock space, which
we choose in the following way: we decompose the gauge field (in momentum space) in terms
of creation and annihilation operators
\be
\label{21}
A (k) = \frac{1}{\sqrt{2 \omega(k)}} \lk a (k) + a^\dagger (- k) \rk ,
\ee
where $\omega (k)$ is an arbitrary (positive definite) kernel. Defining the vacuum state $| k = 0 \rangle$ by
\be
\label{22} 
a (k) | 0 \rangle = 0 
\ee
this state becomes in coordinate representation 
\be
\label{23}
\langle A | 0 \rangle = \exp \lk - \frac{1}{2} \int A \omega A \rk .
\ee
A complete basis of the gluonic Fock space is then given by
\be\label{24}
| 0 \rangle , \quad a^\dagger (k) | 0 \rangle , \quad
a^\dagger (k) a^\dagger (k') | 0 \rangle , \quad\dots
\ee
The exact density matrix (\ref{18}) is too difficult to handle given the complicated
form of the Yang--Mills Hamiltonian. For this purpose we replace the Yang--Mills
Hamiltonian in the density matrix by a single-particle Hamiltonian
\be
\label{25}
D = \exp \bigl[- h/(k_{\textsc{b}} T) \bigr], \qquad h = \int \dbar{k} \: \Omega(k) \, a^\dagger(k) \, a (k) , \qquad 
\dbar{k}\equiv \mathrm{d}^3k/(2\pi)^3.
\ee
Since $h$ is a single-particle Hamiltonian the thermal averages (\ref{19}) with the density
matrix (\ref{25}) can be worked out by using Wick's theorem. For the gluonic occupation
numbers one finds
\be
\label{26}
\langle a (\vk) a^\dagger(\vk') \rangle_T = (2\pi)^3 \delta(\vk-\vk') n(k), \qquad\text{with}\qquad
n (k) = \bigl[ \exp \bigl(\Omega (k)/(k_{\textsc{b}} T)\bigr) - 1 \bigr]^{- 1} ,
\ee
which are the usual Bose occupation numbers with $\Omega (k)$ representing the
single-particle energies. By means of Wick's theorem all thermal averages can then be
expressed in terms of the gluon propagator
\be
\label{27}
\langle A (\vk) A (\vk') \rangle_T = (2\pi)^3 \delta(\vk+\vk') \bigl(1 + 2n(k)\bigr) / \bigl(2 \omega (k)\bigr)  .
\ee
From the density matrix (\ref{25}) we find the entropy $S$ and the free energy $F$
\be\label{29}
S = - k_{\textsc{b}} \Tr D \ln D , \qquad
F = \langle H \rangle_T - TS .
\ee
So far the two kernels $\Omega (k)$, entering the density matrix (\ref{25}), and $\omega (k)$,
entering our basis states (\ref{23}), are completely arbitrary. We now determine $\Omega (k)$
by minimizing the free energy (\ref{29}). Instead of taking the variation with respect
to $\Omega (k)$ it is more convenient to take the variation with respect to the finite
temperature occupation numbers $n (k)$ (\ref{26}), which is equivalent since $n (k)$ is
a monotonic function of $\Omega (k)$. This yields
\be
\label{30}
\Omega (k) = \frac{\delta e [n, \omega]}{\delta n (k)} \, ,
\ee
where $e [n, \omega] = \langle H \rangle_T /\bigl( 2 (N^2_{\mathrm{c}} - 1) \cdot V\bigr)$
is the energy density per gluonic degree of freedom ($V$ is the spatial volume).
From the analogy with Landau's liquid Fermi theory we identify $\Omega (k)$ as a quasi-gluon
energy. Evaluating the thermal expectation value of the Yang--Mills Hamiltonian $\langle H \rangle_T$
up to two loops one finds for the quasi-gluon energy
\be\label{31}
\frac{\Omega (k)}{\omega (k)} =  1 +
\frac{g^2 N_\mathrm{c}}{4} \int \dbar{q} \: F(\vk - \vq) \: \frac{1 + (\hat{k} \cdot \hat{q})^2}{\omega (q)} \: \bigl[1 + 2n (q)\bigr] .
\ee

\begin{figure}
\centering
\parbox[][13ex][b]{.3\linewidth}{%
\centering\includegraphics[width=0.5\linewidth]{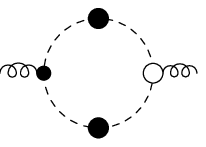}\\(a)}
\hfil
\parbox[][13ex][b]{.3\linewidth}{%
\centering\includegraphics[width=0.5\linewidth]{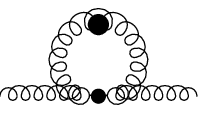}\\[2ex](b)}
\hfil
\parbox[][13ex][b]{.3\linewidth}{%
\centering\includegraphics[width=0.5\linewidth]{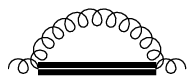}\\[2ex](c)}
\caption{Diagrammatic representation of the contributions to the gap equation: (a) full
ghost loop, (b) tadpole, (c) one-gluon loop contribution from the Coulomb term. Small
black dots and open circles represent, respectively, bare and full vertices of the
Hamiltonian; the double line denotes the Coulomb kernel.}
\label{figure4-1}
\end{figure}

\begin{figure}
\centering\includegraphics[width=.55\linewidth]{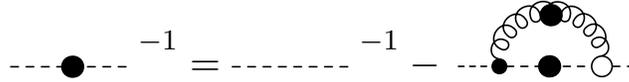}
\caption{\label{figure5}Diagrammatic representation of the ghost DSE.}
\end{figure}

The kernel $\omega (k)$ can be chosen, in principle, completely arbitrary (except that
it has to be positive definite) and the results of our variational calculation should not
depend on the choice of $\omega (k)$. However, since we have introduced approximations
(calculating the energy up to two loops) our results do depend on $\omega (k)$. The best
we can do is to vary the free energy $F$ Eq.~(\ref{29}) with respect to $\omega (k)$. This
guarantees that the free energy is at least to first order independent of $\omega (k)$.
From $\delta F / \delta \omega (k) = 0$ we find the gap equation
\be
\label{32}
\omega^2 (k) = k^2 + \chi^2 (k) + I_\text{tad} [n] + I_\text{c} [n] (k)  ,
\ee
where $\chi (k)$ is the ghost loop, $I_\text{tad} [n]$ is the tadpole and $I_\text{c} [n] (k)$
the one gluon-loop contribution from the Coulomb term, see Fig.~\ref{figure4-1}.
The finite-temperature modifications arise exclusively from the finite-temperature part
of the gluon propagator, Eq.~(\ref{27}), which depends on the occupation number $n (k)$.
The finite-temperature modifications are all ultraviolet finite so the renormalization
of the gap and Dyson--Schwinger equations can be done in exactly the same way as at zero
temperature, see Ref.~\cite{Reinhardt:2007wh}. The gap equation has to be solved together
with the Dyson--Schwinger equation for the ghost propagator, which is illustrated in 
Fig.~\ref{figure5}. This equation is the same one as at zero temperature except that
the gluon propagator is replaced by its finite-temperature counterpart (\ref{27}).
As we have illustrated above for zero temperature, the Coulomb term of the Yang--Mills
Hamiltonian does barely influence the ghost and gluon propagator. Therefore we will
ignore the Coulomb term in the following.\footnote{We should, however, stress that the
Coulomb term is utterly important for the quark sector since it provides the confining
potential for the quarks, see further below.} Neglecting the Coulomb term leads to
substantial simplifications \cite{R14}. The quasi-gluon energy $\Omega (k)$ of the
density matrix (\ref{25}) and the kernel $\omega (k)$ of the vacuum wave functional
(\ref{23}) become then identical. Furthermore, the ghost Dyson--Schwinger equations and
the gap equation then decouple from the Dyson-Schwinger equation for the Coulomb form
factor \cite{Feuchter:2004mk}.

%%%%%%%%%%%%%%%%%%%%%%%%%%%%%%%%%%%%%%%%%%%%%%%%%%%%%%%%%%%%%%%%%%%%%%%%%%%%%%%%%%%%%%%%%
%%%%%%%%%%%%%%%%%%%%%%%%%%%%%%%%%%%%%%%%%%%%%%%%%%%%%%%%%%%%%%%%%%%%%%%%%%%%%%%%%%%%%%%%%

\section{Infrared analysis}

Before we present the numerical results let us summarize the results of an infrared
analysis for the remaining gap and ghost Dyson-Schwinger equations. Let us first
recapitulate the result of the infrared analysis at zero temperatures \cite{Schleifenbaum:2006bq}.

For the infrared analysis we assume power-law behaviours of the propagators involved
\be
\label{35}
\omega (p\to0) = A /p^\alpha , \qquad d (p\to0) = B /p^\beta .
\ee
Assuming the horizon condition $d^{- 1} (p = 0) = 0$, which implies $\beta > 0$, we
find from the ghost DSE the sum rule
\be
\label{36}
\alpha = 2 \beta + 2 - d ,
\ee
where $d$ is the number of spatial dimensions. For Coulomb gauge in $d = 3$ we obtain $\alpha = 2 \beta - 1$,
and including also the gap equation one finds the following two solutions
\be
\label{37}
\beta = 1 \; (1.001) , \qquad \beta = 0.796 \;  (0.794) ,
\ee
where the results from the numerical evaluation are given in the bracket. In $d = 2$
dimensions one finds a single solution, which within the angular approximation is given by
\be
\label{38}
\beta = 0.5 \; (0.45) 
\ee
while without the angular approximation one finds $\beta= 0.4$.

At arbitrary finite temperature the infrared analysis cannot be carried out since the
gluon energy $\Omega (k) = \omega (k)$ enters the thermal occupation numbers $n (k)$,
Eq.~(\ref{26}), in the exponent. However, at very high temperatures we can expand this
exponent and the occupation numbers reduce to
\be
\label{39}
n (k) = k_{\textsc{b}} T / \omega (k) .
\ee
Then $\omega (k)$ enters only as power in the gap and DSE and we can carry out the
infrared analysis in the standard fashion. One finds the same sum rule Eq.~(\ref{36}) as
at zero temperature. However, one obtains only a single solution for the infrared exponent 
\be
\label{40}
\beta = 1 / 2 , \qquad \alpha = 0 .
\ee
This is the infrared exponent for the ghost in $d = 2$ dimensions, which reflects the fact 
that at high temperature a dimensional reduction to the $2 + 1$-dimensional theory occurs. 

%%%%%%%%%%%%%%%%%%%%%%%%%%%%%%%%%%%%%%%%%%%%%%%%%%%%%%%%%%%%%%%%%%%%%%%%%%%%%%%%%%%%%%%%%
%%%%%%%%%%%%%%%%%%%%%%%%%%%%%%%%%%%%%%%%%%%%%%%%%%%%%%%%%%%%%%%%%%%%%%%%%%%%%%%%%%%%%%%%%

\section{Numerical results}

\begin{figure}[t]
\centering
\begin{minipage}[t]{.45\linewidth}
\includegraphics[width=\linewidth]{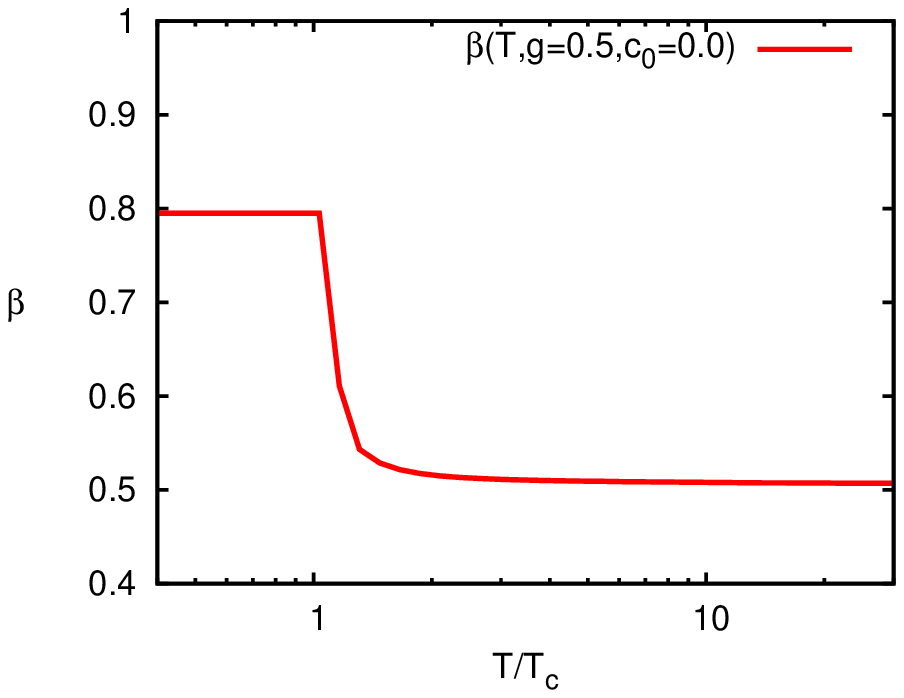}
\caption{IR exponent of ghost as function of the temperature.}
\label{figure6}
\end{minipage}
\hfil
\begin{minipage}[t]{.45\linewidth}
\includegraphics[width=\linewidth]{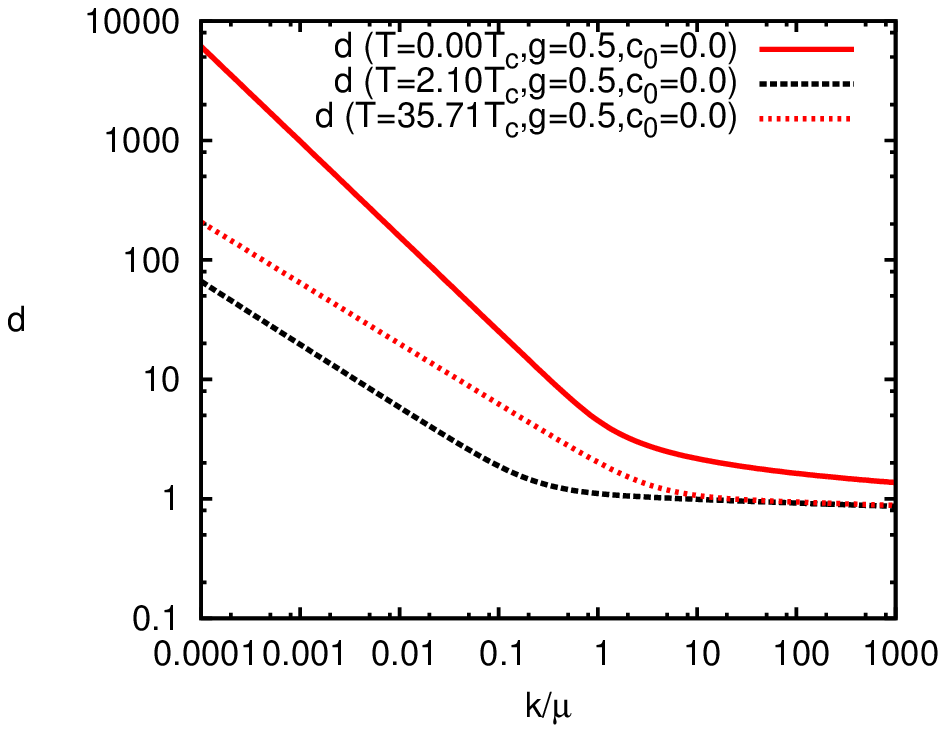}
\caption{The ghost form factor.}
\label{figure7}
\end{minipage}
\end{figure}

\begin{figure}[t]
\centering
\parbox{.48\linewidth}{
\includegraphics[width=\linewidth]{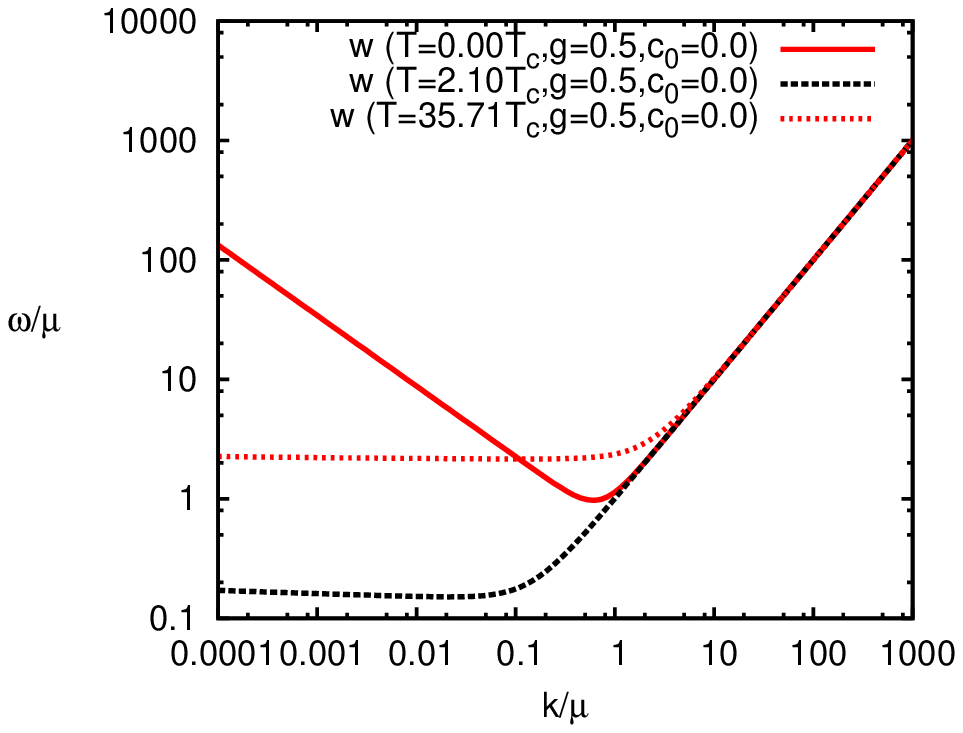}
\caption{The gluon energy.}
\label{figure8}
}
\hfil
\parbox{.48\linewidth}{
\includegraphics[width=\linewidth]{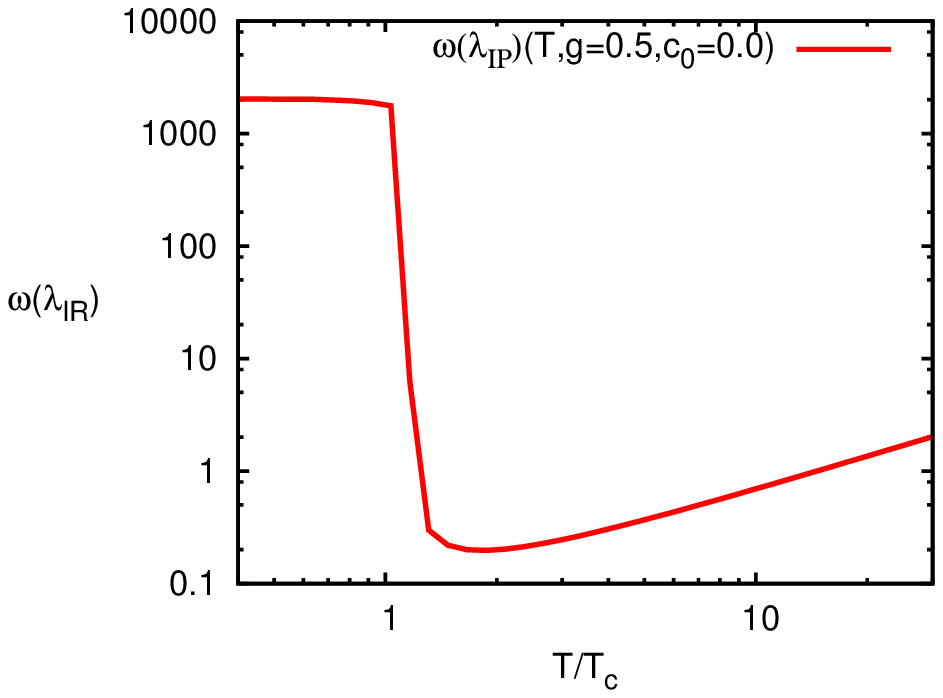}
\caption{The IR mass of the gluon.}
\label{figure9}
}
\end{figure}

Figure \ref{figure6} shows the ghost infrared exponent $\beta$ determined from the
numerical solutions as function of the temperature. As one observes the infrared exponent
stays constant below a critical temperature where it suddenly drops and approaches the
value $\beta = 1/2$ for high temperatures, in agreement with the infrared analysis.
Both zero-temperature solutions,
$\beta = 1$ and $\beta = 0.796$, convert to the same high-temperature solution.
Figures \ref{figure7} and \ref{figure8} show the numerical solutions for the ghost form
factor and the gluon energy for zero temperature and above the deconfinement temperature.
One observes that the infrared behaviour is indeed in accord with the findings of the infrared
analysis.
The sudden drop of the infrared exponent of the ghost form factor (see Fig.~\ref{figure6})
can be used to define the deconfinement phase transition temperature $T_c$. Fitting at
$T=0$ the numerical solution for $\omega(k)$ to the lattice gluon propagator \cite{Burgio:2008jr}
to fix the physical scale one finds for the critical temperature of the deconfinement
phase transition
\be
\label{48}
T_c \sim 270 \dots 290 \mbox{ Mev} ,
\ee
for $SU(2)$, which compares well with the lattice result of $T_c = 295$ MeV.

\begin{figure}[t]
\centering
\parbox[t]{.45\linewidth}{
\includegraphics[width=\linewidth]{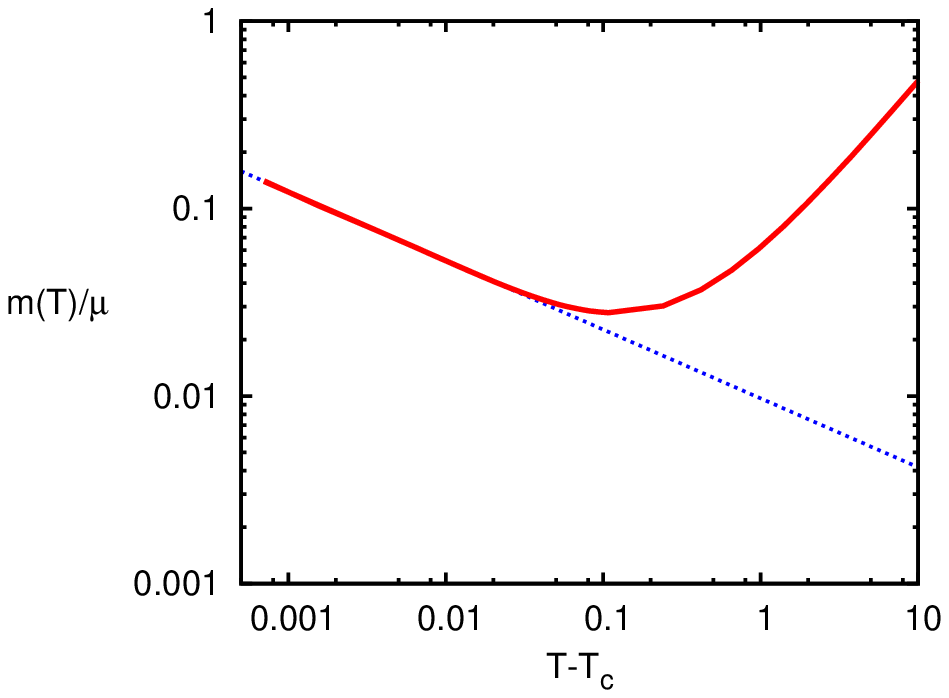}
\caption{Critical behaviour of the effective gluon mass.}
\label{figure10}
}
\hfil
\centering
\parbox[t]{.45\linewidth}{
\includegraphics[width=\linewidth]{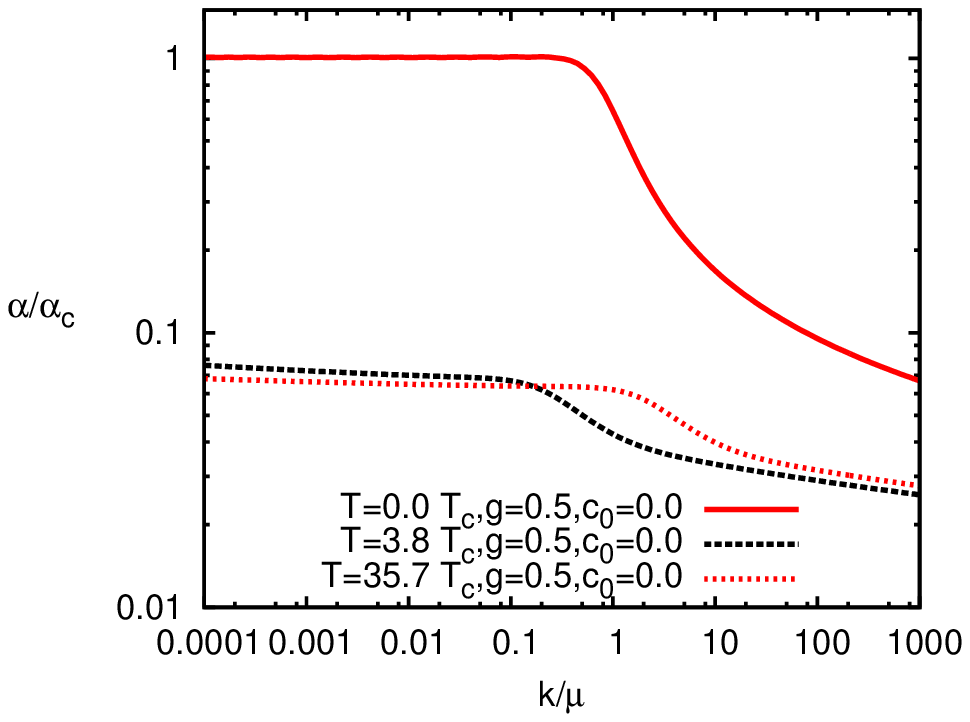}
\caption{Running coupling constant from the ghost-gluon vertex.}
\label{figure11}
}
\end{figure}

Figure \ref{figure9} shows the infrared mass of the gluon defined by the gluon energy at
the numerical infrared scale $\lambda_{\mathrm{IR}}$,
\be
\label{45}
m (T) = \omega (k = \lambda_{\mathrm{IR}}, T) .
\ee
This mass behaves similarly as the infrared exponent of the ghost form factor. It stays
constant below a critical temperature, where it suddenly drops and after that starts
rising linearly with the temperature. 
Zooming into the behaviour of $m (T)$ in the transition regime of Fig.~\ref{figure9},
see Fig.~\ref{figure10}, one finds for the critical exponent of the effective gluon mass
defined by
\be
\label{50}
m (T) \sim \lk T /T_c - 1 \rk^{- \kappa} 
\ee
a value of $\kappa \simeq 0.37$. This compares well with the result of $\kappa = 0.41$
obtained in Ref.~\cite{R17}, where a quasi-gluon picture has been used to fit the lattice
results for the energy density and the pressure, and assuming
furthermore input from the $d = 3$ Ising model, which is in the same universality class
as $SU(2)$ gauge theory.

Figure \ref{figure11} shows the running coupling constant at zero temperature and above
the deconfinement phase transition. One observes a substantial reduction of the low
energy plateau.

%%%%%%%%%%%%%%%%%%%%%%%%%%%%%%%%%%%%%%%%%%%%%%%%%%%%%%%%%%%%%%%%%%%%%%%%%%%%%%%%%%%%%%%%%
%%%%%%%%%%%%%%%%%%%%%%%%%%%%%%%%%%%%%%%%%%%%%%%%%%%%%%%%%%%%%%%%%%%%%%%%%%%%%%%%%%%%%%%%%

\section{Hamiltonian approch to QCD}

\begin{figure}[t]
\begin{minipage}[t]{.48\linewidth}
\includegraphics[width=\linewidth]{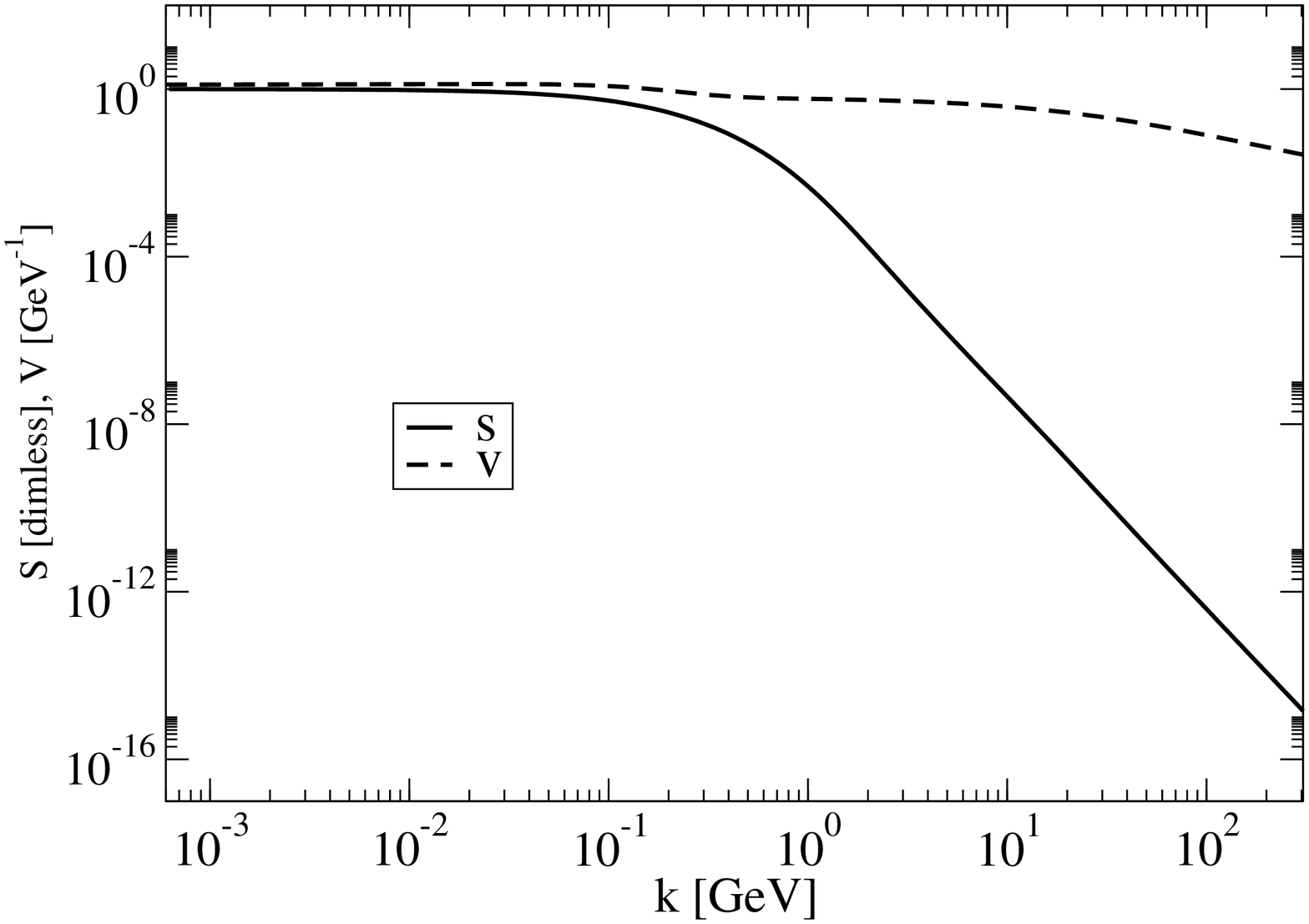}
\caption{The variational kernels $S$ and $V$.}
\label{fig-s-v-asymptotic}
\end{minipage}
\hfil
\begin{minipage}[t]{.48\linewidth}
\includegraphics[width=\linewidth]{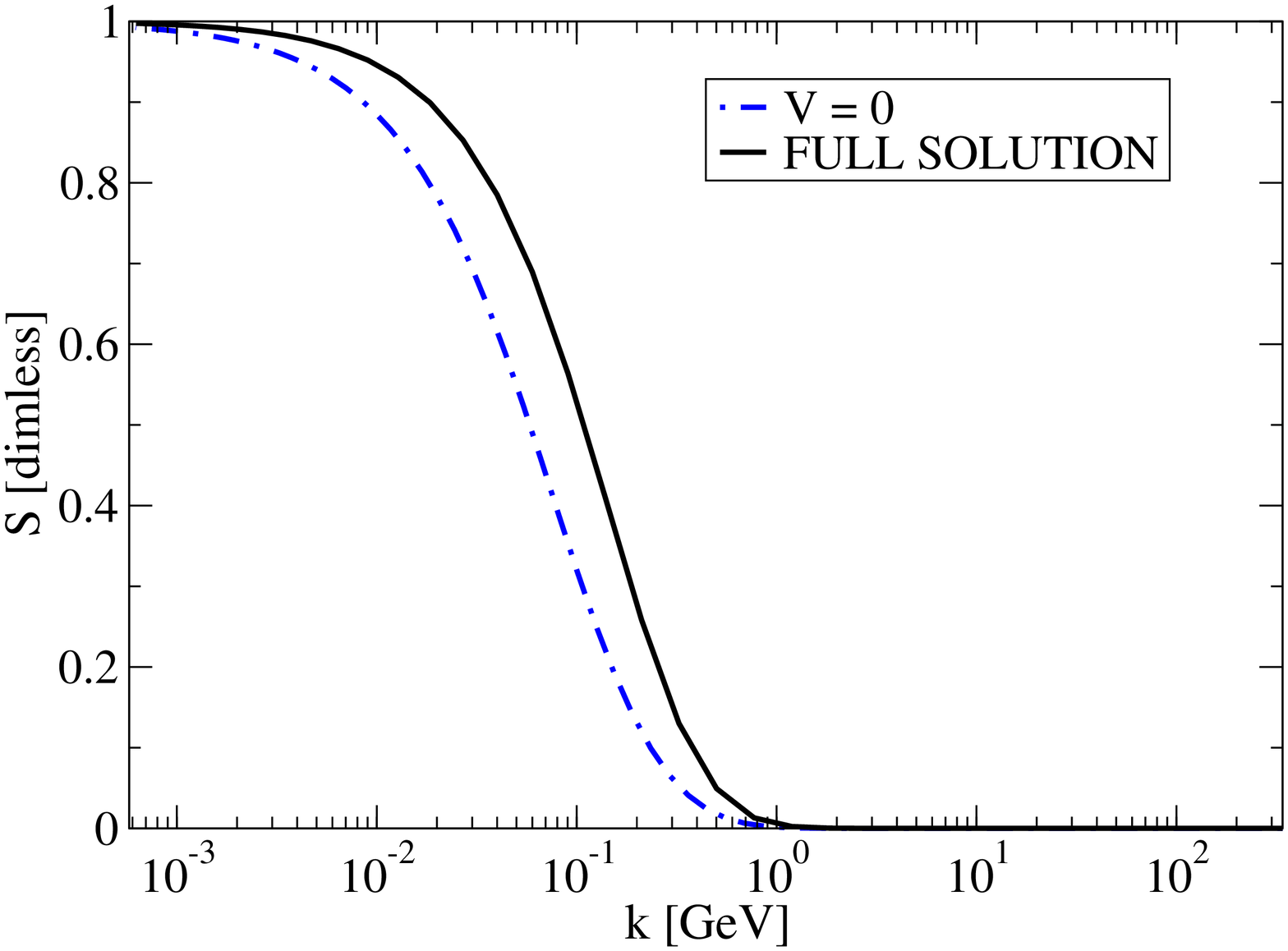}
\caption{Comparison of scalar kernel $S(k)$ with ($V \neq 0$) and without $(V = 0)$ the
quark-gluon coupling included.}
\label{figure12}
\end{minipage}
\end{figure}

When the quarks are included the Hamiltonian (\ref{4}) has to be supplemented by the quark term
\be
\label{52}
H_q = \int q^\dagger(\vx) \left[ \vec{\alpha} \lk \vp + g \vA \rk + \beta m_0 \right] q(\vx) ,
\ee
where $q(\vx)$ denotes the quark field operator and $m_0$ is the current quark mass.
Furthermore, the matter charge density in the Cou\-lomb Hamiltonian (\ref{6}) becomes
$\rho^a_m (x) = q^\dagger (x) t^a q(x)$, where $t^a$ are the generators of the gauge
group in the fundamental representation. For the quark sector we use the following
ansatz for the vacuum wave functional \cite{Pak:2011wu}
\be
\label{53}
| \phi \rangle^{}_{\textsc{q}} = \exp \left[ \int q^\dagger \lk S \beta + V \vec{\alpha} \cdot \vA \rk q \right] | 0 \rangle ,
\ee
where $| 0 \rangle$ is the perturbative quark vacuum state, which describes a system of
free quarks with mass $m_0$. Furthermore $S$ and $V$ are variational kernels to be determined
by minimizing the energy density. For $V = 0$ Eq.~(\ref{53}) defines a BCS-type wave
functional considered in Ref.~\cite{Adler:1984ri}. The new important aspect of the wave
function (\ref{53}) is the coupling of the quarks to the gauge field with form factor $V$.
Without this term the quark-gluon coupling in the quark Hamiltonian (\ref{52}) escapes
the expectation value. The total QCD wave functional is then given by
\be
\label{54}
| \Phi \rangle = | \psi \rangle^{}_{\textsc{ym}} \otimes | \phi \rangle^{}_{\textsc{q}} \, ,
\ee
where $| \psi \rangle_{\textsc{ym}}$ is the Yang--Mills vacuum wave functional
given in Eq.~(\ref{8}). We keep $\omega$ fixed at its form determined from the Yang-Mills sector.
We minimize the quark energy
$\langle \langle H_q \rangle_\textsc{q}\rangle_{\textsc{ym}}$ with respect to the kernels 
$S$ and $V$. The result of this variation is shown in Fig.~\ref{fig-s-v-asymptotic}.
Figure \ref{figure12} compares the scalar form factor with and without taking into account the quark-gluon coupling. 
With the quark-gluon coupling switched off ($V = 0$) we find the value
for the quark condensate 
\be
\label{57}
\langle \bar{q} q \rangle = \Bigl( - 113\mbox{ MeV} \sqrt{\sigma_{\textsc{c}} / \sigma_{\textsc{w}}} \Bigr)^3 ,
\ee
which corresponds to the result of Ref.~\cite{Adler:1984ri}, while with the quark gluon
coupling included ($V \neq 0$) we obtain 
\be
\label{58}
\langle \bar{q} q \rangle = \lk - 135\mbox{ MeV} \sqrt{\sigma_{\textsc{c}} / \sigma_{\textsc{w}}} \rk^3 .
\ee
This is an increase of the figure in the bracket by about 20\%. In Eqs.~\eqref{57} and
\eqref{58} $\sigma_{\textsc{c}}$ and $\sigma_{\textsc{w}}$ denote respectively the
Coulomb and Wilsonian string tension. Lattice results show that the ratio of these
quantities is given in the range of
\be
\label{47}
\sigma_{\textsc{c}} / \sigma_{\textsc{w}} \simeq 2 \dots 3 .
\ee
This implies a quark condensate in the range of 
\be
\label{59}
\langle \bar{q} q \rangle = ( - 191 \dots 234\mbox{ MeV} )^3 ,
\ee
which compares well with the phenomenological value of 
\be
\label{60}
\langle \bar{q} q \rangle = ( - 230\mbox{ MeV} )^3 .
\ee

\begin{figure}[t]
\centering
\includegraphics*[width=0.45\linewidth]{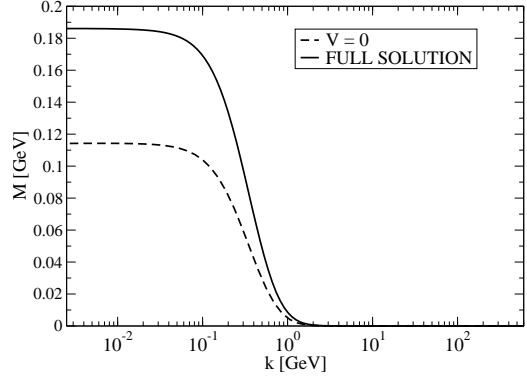}
\caption{Running quark mass.}
\label{fig-M}
\end{figure}
Figure \ref{fig-M} shows the dynamical quark mass as a function of the momentum with and
without the quark gluon coupling. We observe an essential increase in the quark mass
 when the coupling of the quarks to the gluons is included. For the constituent quark
mass defined by the zero momentum value of the running quark mass $M = M(k = 0)$ we obtain 
without the quark-gluon coupling ($V = 0$) 
\be
\label{62}
M = 84\mbox{ MeV} \sqrt{\sigma_{\textsc{c}} / \sigma_{\textsc{w}}}
\ee
and with the quark-gluon coupling included ($V \neq 0$) 
\be
\label{63}
M = 132\mbox{ MeV} \sqrt{\sigma_{\textsc{c}} / \sigma_{\textsc{w}}} \, ,
\ee
which is an increase by 57\%. Assuming again the lattice result Eq.~\eqref{47} for the
ratio of the string tensions we obtain \cite{Pak:2011wu} constituent masses in the range of
\be
\label{64}
M = 187 \dots 230\mbox{ MeV} ,
\ee
which brings the constituent mass into the region of its phenomenological value. 

The results obtained so far in the variational approach to QCD in Coulomb gauge are very
encouraging and call for more detailed investigations.

%%%%%%%%%%%%%%%%%%%%%%%%%%%%%%%%%%%%%%%%%%%%%%%%%%%%%%%%%%%%%%%%%%%%%%%%%%%%%%%%%%%%%%%%%
%%%%%%%%%%%%%%%%%%%%%%%%%%%%%%%%%%%%%%%%%%%%%%%%%%%%%%%%%%%%%%%%%%%%%%%%%%%%%%%%%%%%%%%%%

\section*{Acknowledgements}
\vspace*{-2ex}\noindent
It is a pleasure to thank A.~Szczepaniak for collaboration and discussions on parts of the
subjects discussed in this talk.
This work was supported by the Deutsche Forschungsgemeinschaft (DFG) under contract DFG-Re856/6-3,
by the Europ\"aisches Graduiertenkolleg ``Hadronen im Vakuum, Kernen und Sternen'', and
by the Graduiertenkolleg ``Kepler-Kolleg: Particles, Fields and Messengers of the Universe''.

\end{document}